\documentclass[a4paper,11pt]{article}
\usepackage{pos}

\usepackage{natbib}
\title{Calibration and testing of the JEM--EUSO detectors using stars observed in the UV band}
 \ShortTitle{Calibration of the EUSO-TA with stars}

\author*[a,b]{Zbigniew Plebaniak}
\author[c]{Marika Przybylak}

\affiliation[a]{Universit\`{a} di Roma Tor Vergata - Dipartimento di Fisica\\
  Via della Ricerca Scientifica 1 – 00133, Roma, Italy}

\affiliation[b]{Istituto Nazionale di Fisica Nucleare - Sezione di Roma Tor Vergata\\
Via della Ricerca Scientifica 1 – 00133, Roma, Italy}

\affiliation[c]{National Centre for Nuclear Research, Otwock, Poland}

\onbehalf{for the JEM-EUSO Collaboration\\[-1mm]{\normalsize \normalfont (a complete list of authors can be found at the end of the proceedings)}}

\emailAdd{zbigniew.plebaniak@roma2.infn.it}

\abstract{
The JEM-EUSO program is focused on observations of Ultra High Energy Cosmic Rays (UHECRs) from space. 
For this purpose, a series of detectors based on multi-anode photomultiplier tubes with a time resolution of the order of $\mu$s have been developed.
The detectors work in the UV band to search for ultra-fast signals produced in the Earth's atmosphere during an Extensive Air Shower (EAS) development.
Since 2014, various signals have been detected by ground-, ballon- and space-based detectors.
A single photodetector module consists of a focal surface with a matrix of 36 multi-anode photomultiplier tubes containing 2304 pixels.
The detector's structure allows probing it during the mission if a point-like source emitting in a UV band is in the field of view. 
In this work, we present the idea and results of calibration of the JEM--EUSO detectors using signals from stars registered during sky observations from the ground.
Registered signals can be used for the absolute calibration of the detectors and for testing the detector condition during observations.
The presented analysis is based on the data taken by the EUSO-TA and EUSO-TA2 experiments.
}

\ConferenceLogo{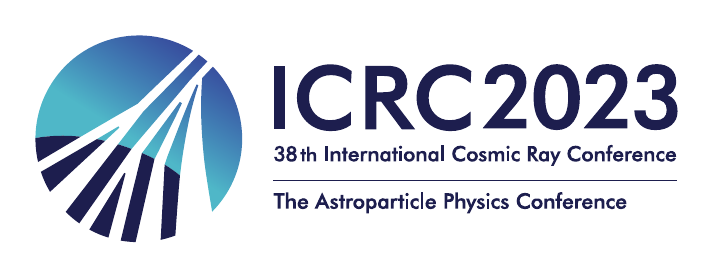}

\FullConference{%
38th International Cosmic Ray Conference (ICRC2023)\\
  26 July - 3 August, 2023\\
  Nagoya, Japan}

\begin{document}
\maketitle

\section{Introduction}
{
The EUSO-TA experiment is a ground-based detector in the JEM-EUSO scientific program. It is a small fluorescence telescope located in Telescope Array (TA)~\cite{TAMEDA200974} site in front of Black Rock Mesa fluorescence station (BRM-TA FD) in the Utah desert. 
The focal surface contains one photodetector module (PDM) composed of 36 Hamamatsu R11265-M64 Multi-Anode Photomultipliers (MAPMTs), resulting in a matrix of 2304 pixels. 
One segment of $2\times2$ MAPMTs creates an Elementary Cell (EC) with a dedicated HVPS module. The detector can observe the night sky to search for Ultra High Energy Cosmic Rays being triggered by TA. 
More details can be found in~\cite{euso-ta-icrc2023}. 
Since its installation in 2013, the detector has proved helpful for testing and developing EUSO detectors and its firmware. 
In 2022, an upgraded version of the so-called EUSO-TA2 was set up at the BRM-TA FD site. 
PDM has been upgraded with a new high voltage unit (HVPS), while the whole optical system in the old design has been conserved. 
The focal surface of the new PDM is presented in figure~\ref{figNewPdm}.

\begin{figure}[ht]
\centering
\includegraphics[width=0.447\textwidth]{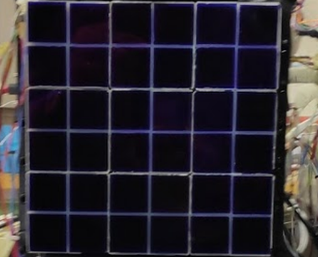}
\hspace{0.5cm}
\includegraphics[width=0.38\textwidth]{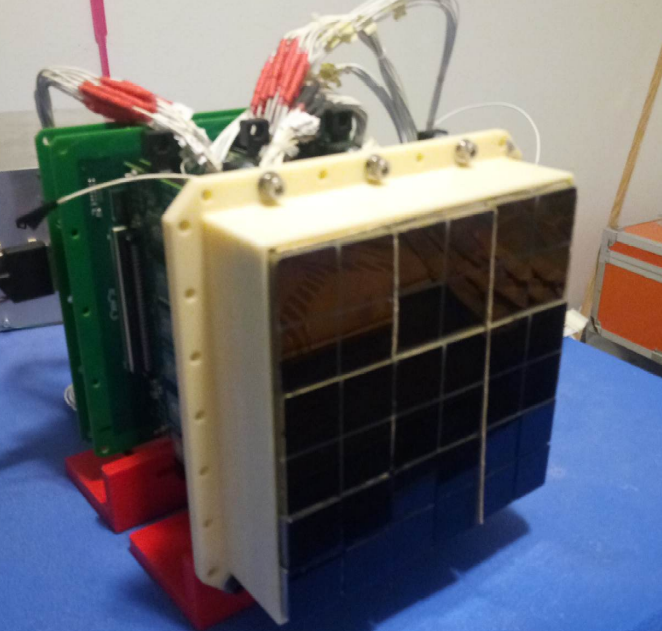}
\caption{Photo-Detector Module of EUSO-TA2 experiment. View from the focal surface (left) and the whole PDM with a visible PDM board at the bottom (right). The focal surface comprises the matrix of 36 MAPMTs divided into 9 ECs. White cables are connected to HVPS and LVPS modules during measurements. The detector is placed at about 43 cm from the first lens.}
\label{figNewPdm}
\end{figure}
}

\begin{figure}[ht]
\centering
\includegraphics[width=0.49\textwidth]{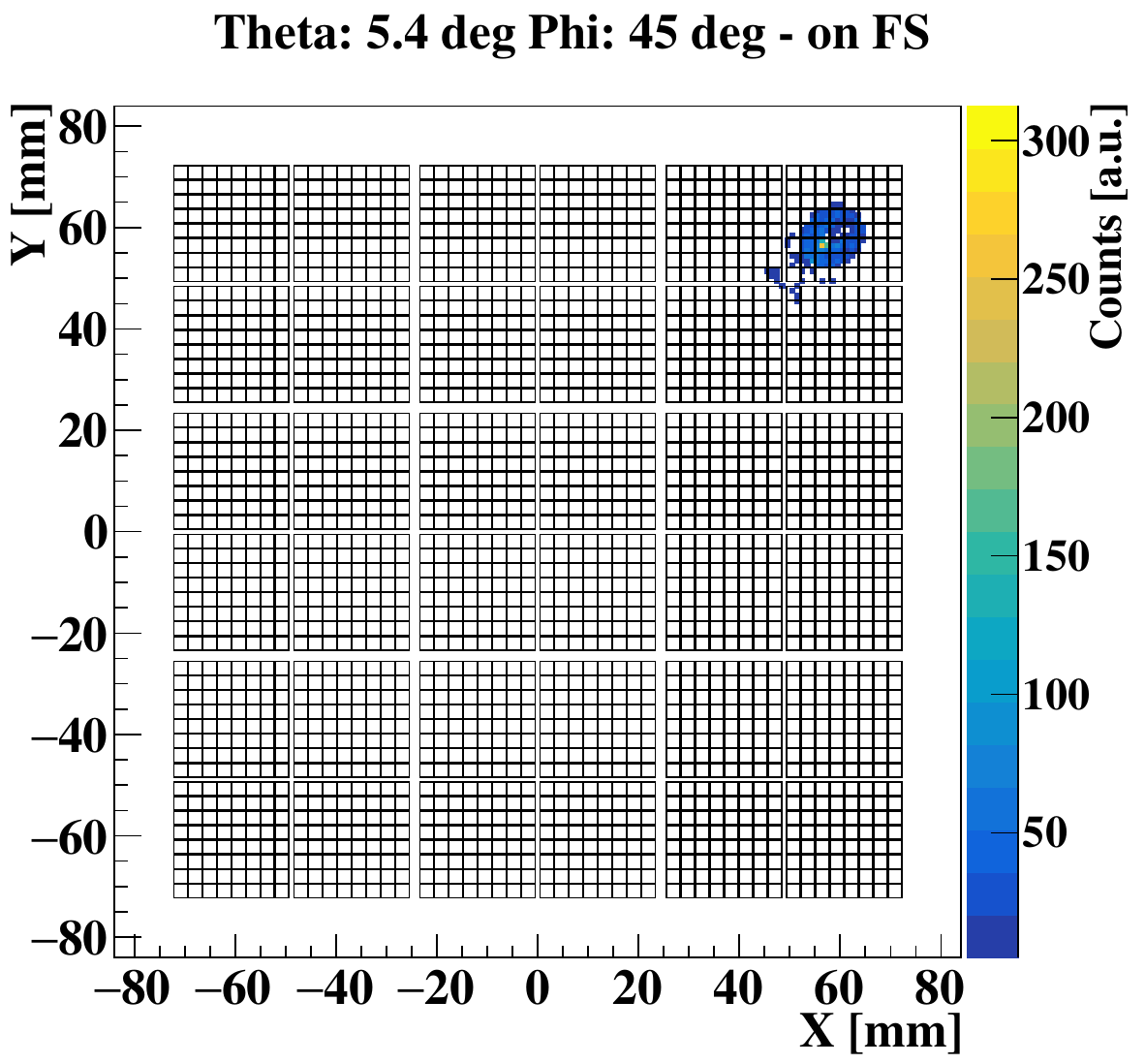}
\includegraphics[width=0.49\textwidth]{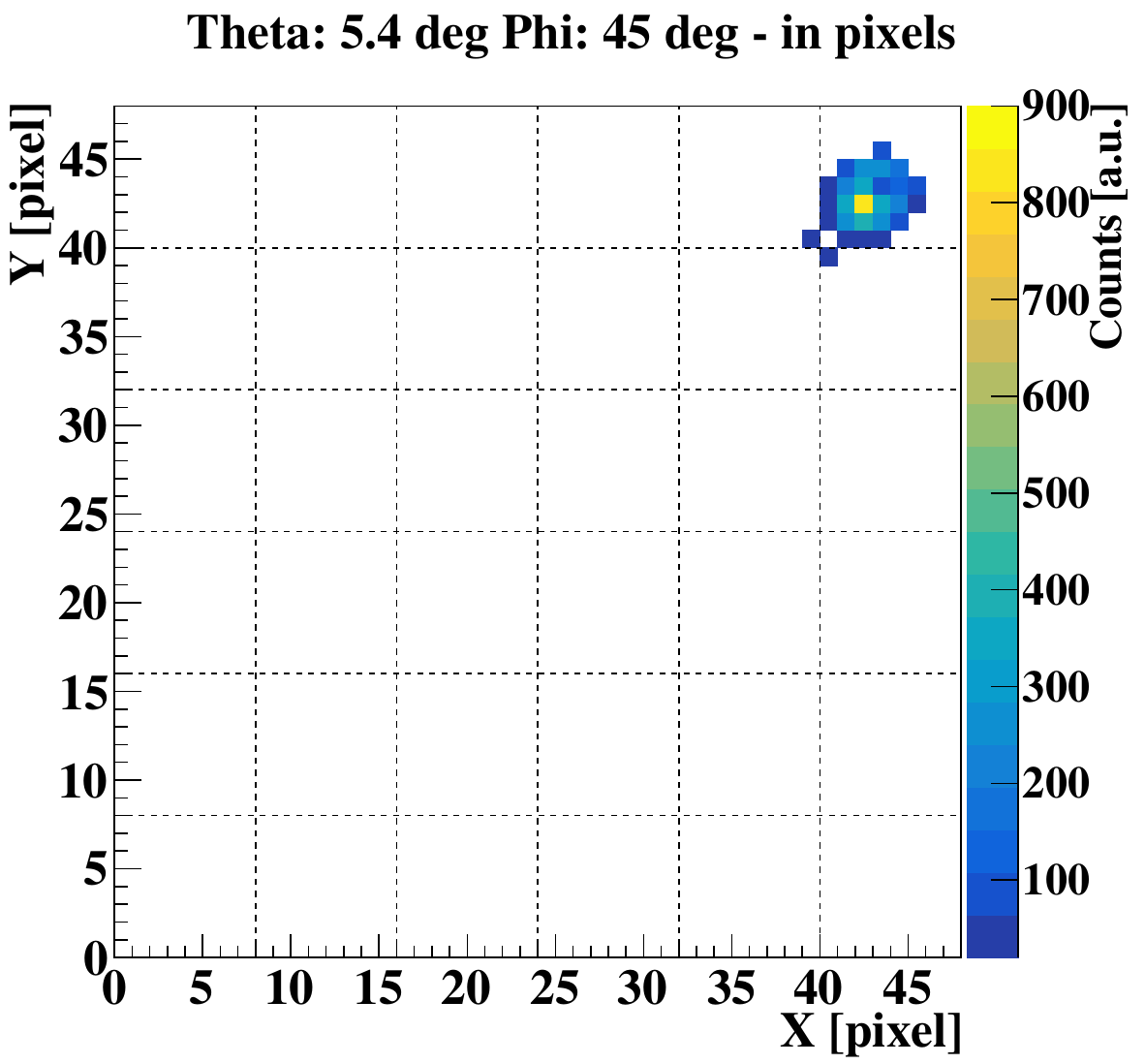}\\
\vspace{0.6cm}
\includegraphics[width=0.24\textwidth]{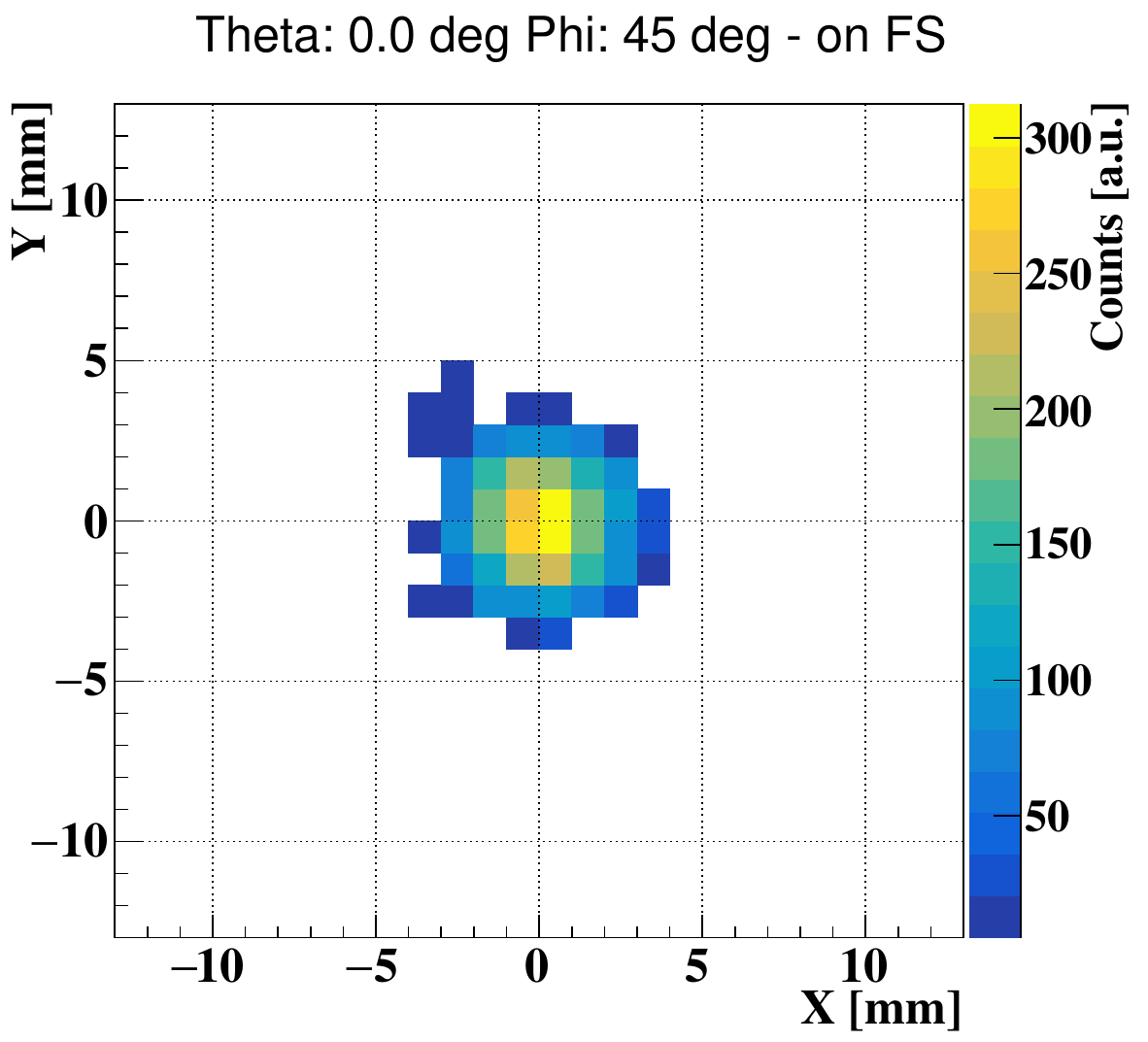}\includegraphics[width=0.24\textwidth]{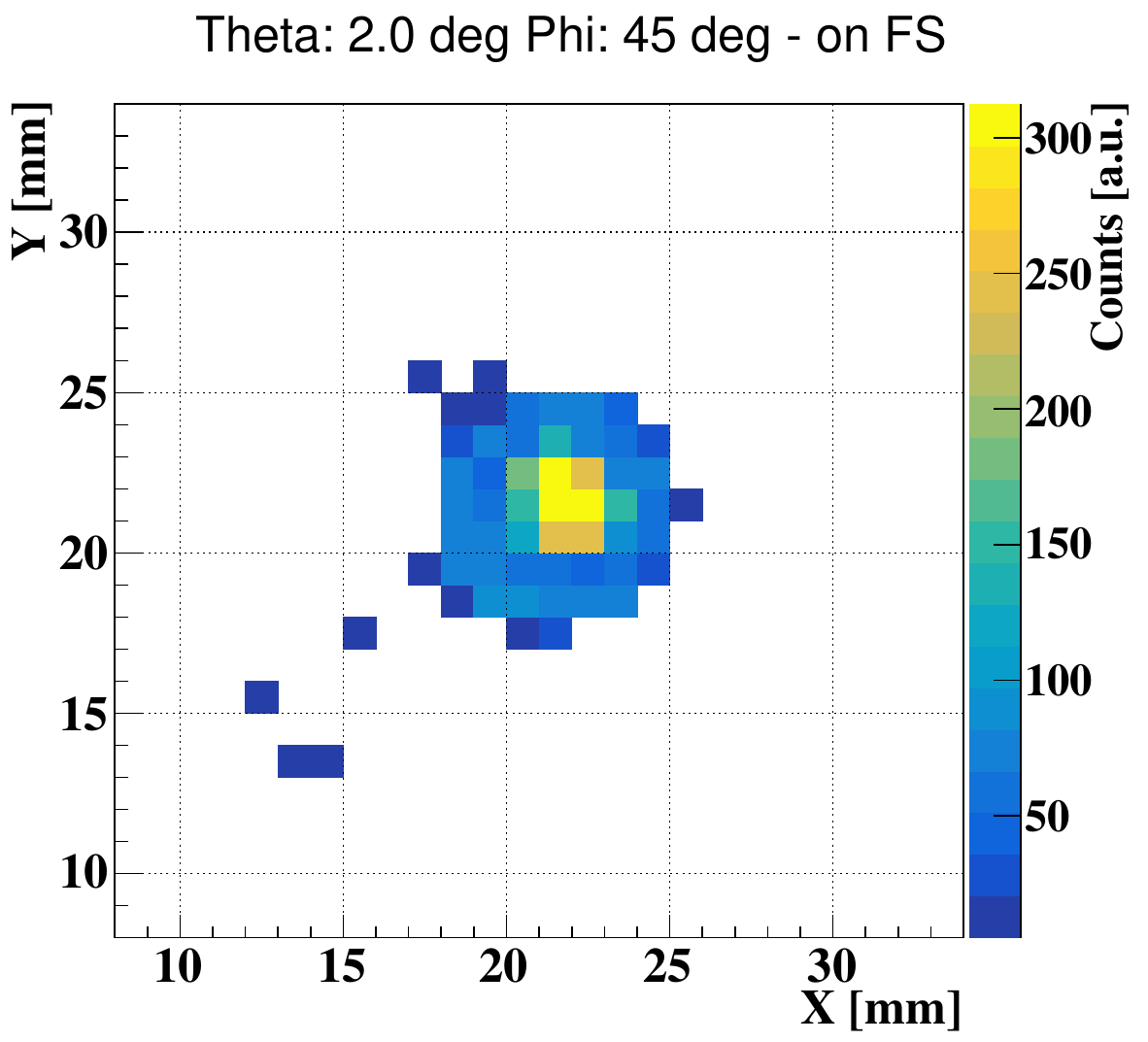}\includegraphics[width=0.24\textwidth]{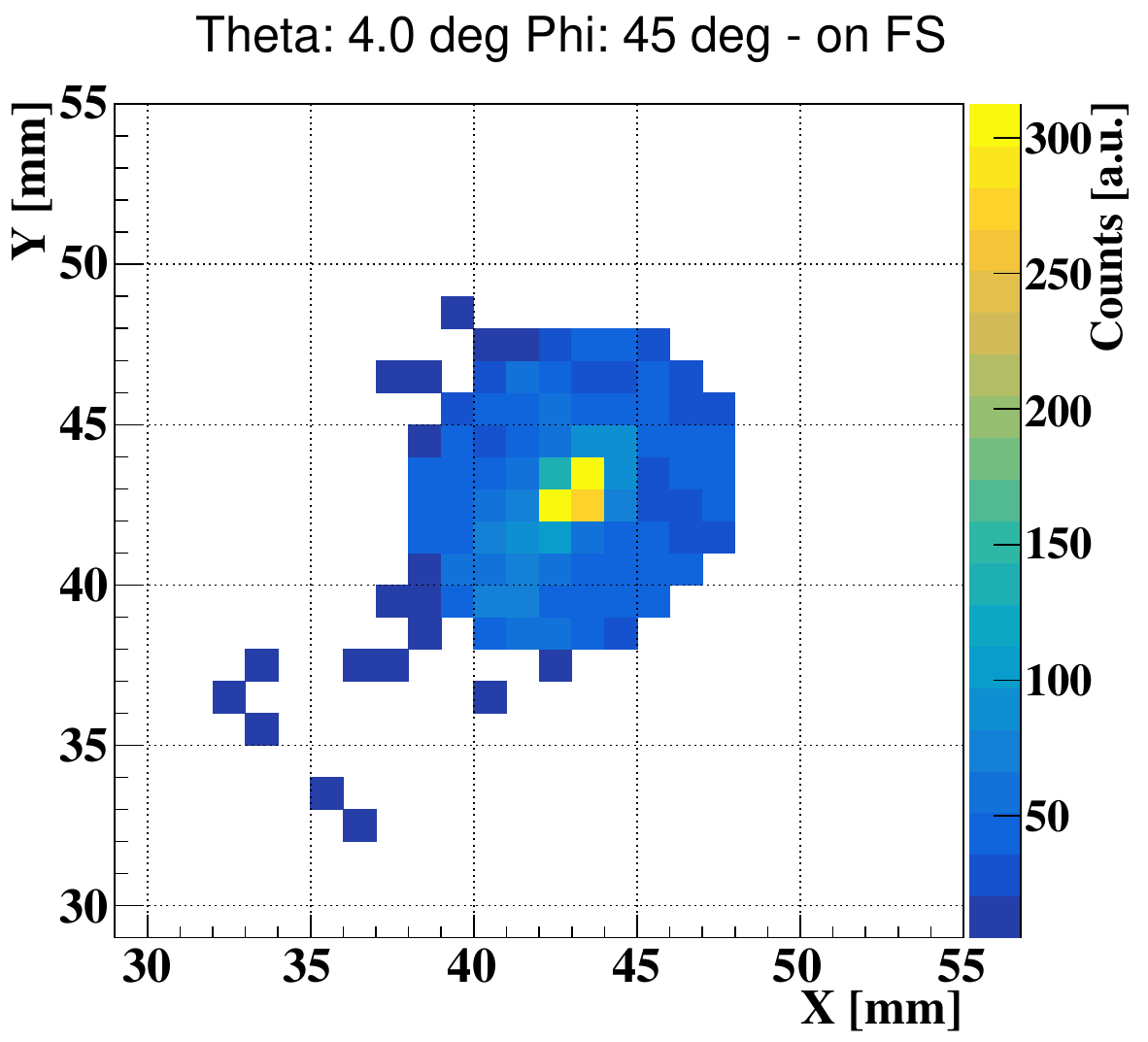}\includegraphics[width=0.24\textwidth]{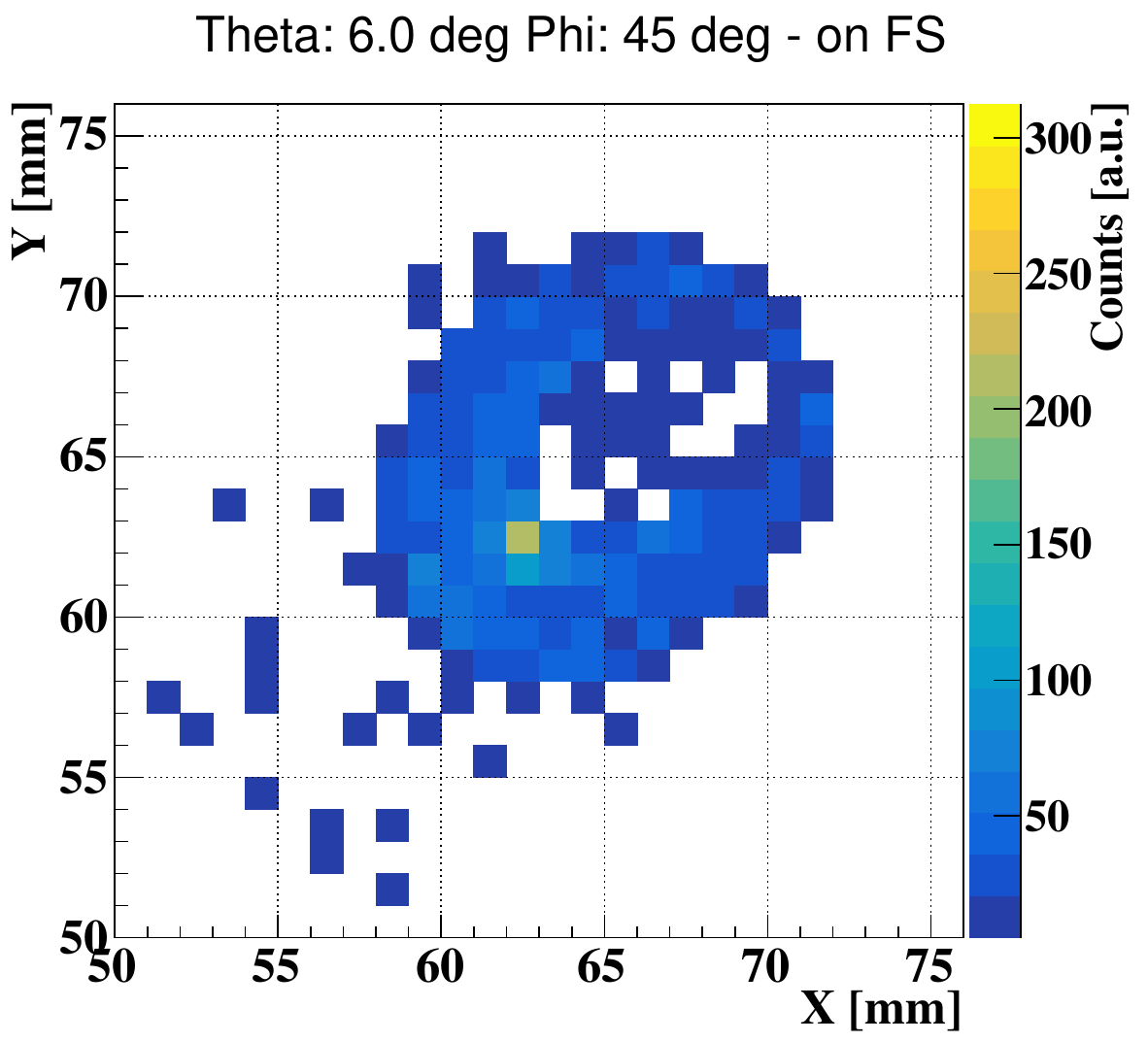}
\caption{EUSO-TA point spread function simulated with OffLine software \cite{JEMEUSOCollaborationPaulBertaina2017_1000082477}. In the left top panel, we show the distribution of simulated photons reaching the focal surface. The simulation includes 100 000 photons at 355 nm hitting the first lens of the detector at 5.4$^\circ$ of incident theta angle. The black grid represents the position of single pixels. Individual PMTs as clusters of 64 pixels and ECs as clusters of 4 PMT can be recognized. The top right plot shows the same result taking into account only the number of counts in individual pixels. Photons hitting the gaps between PMTs or ECs are discarded. Four plots in the bottom panel present simulated distributions of photons on the focal surface for different incident angles: 0$^\circ$, 2$^\circ$, 4$^\circ$ and 6$^\circ$. More detailed analysis shows that even for 6$^\circ$ of incident angle, more than 50$\%$ of the signal should be placed in the area of 3x3 pixels.}
\label{OffLineSimu}
\end{figure}

The EUSO-TA optical system comprises two Fresnel lenses produced from UV-transmitting polymethyl-methacrylate (PMMA) with about 40$\%$ overall efficiency.
The width of the PSF increases with the distance from the optical axis due to optical aberration. 
The EUSO-TA detector, including optics, has been implemented in OffLine~\cite{JEMEUSOCollaborationPaulBertaina2017_1000082477} and ESAF~\cite{Berat:2009va} frameworks to perform related simulations. 
In figure~\ref{OffLineSimu}, we present an example of the PSF simulated with OffLine software to illustrate its dependency on the position on the focal surface. 
A portion of the signal that can be registered in a defined area on the focal surface due to the PSF is one of the critical parameters during absolute calibration with a point-like signal. 
More details regarding EUSO-TA optics design can be found in~\cite{2013ICRC...33..627T} and \cite{Abdellaoui:2018rkw}.

\section{Measurements and data structure}

EUSO-TA detector in the first version was operating in single photons counting mode, and the number of photons is used to be read from each pixel and reset every 2.5$\mu s$, which defines Gate Time Unit (GTU). 
The dead time is about 0.2$\mu s$ in this case.
In 2015--2016, the detector was externally triggered for CR events by the TAFD station with an average trigger rate of about 4~Hz. 
An upgraded version of the PDM in EUSO-TA2 operating since 2022 allows for data acquisition in three-time scales: D1, D2 and D3 with integration times of 2.5~$\mu$s, 320~$\mu$s and 40.96~ms respectively and without an external trigger. 
In this work, we focus on the calibration of the detector using point-like signals emitted by stars in the field of view. We present mainly EUSO-TA data, but to test algorithms, we also use the D3 acquisition mode of EUSO-TA2, which allows us to limit the dead time in a significant way.

\section{Data processing}
{
The data analysis process to extract signals from stars has been described in detail in~\cite{JEM-EUSO:2021lpf}. 
Working on counts summed over 128 GTUs, we were able to fit tracks of stars in FoV with M$_{B}$ < 5.5$^{m}$ for EUSO-TA and M$_{B}$ < 6.4$^{m}$ for EUSO-TA2. 
An example of this fitting is presented in figure~\ref{DataAnalysis}. 
Based on the fit, the position of the signal on the focal surface can be determined with an accuracy of about 0.4 pixels. If the track is close to the centre of a specific pixel,  information about the background and signal level can be extracted from the light curve calculated for a 3x3 pixels area with the centre at the selected pixel.

\begin{figure}[ht]
\centering
\includegraphics[width=0.33\textwidth]{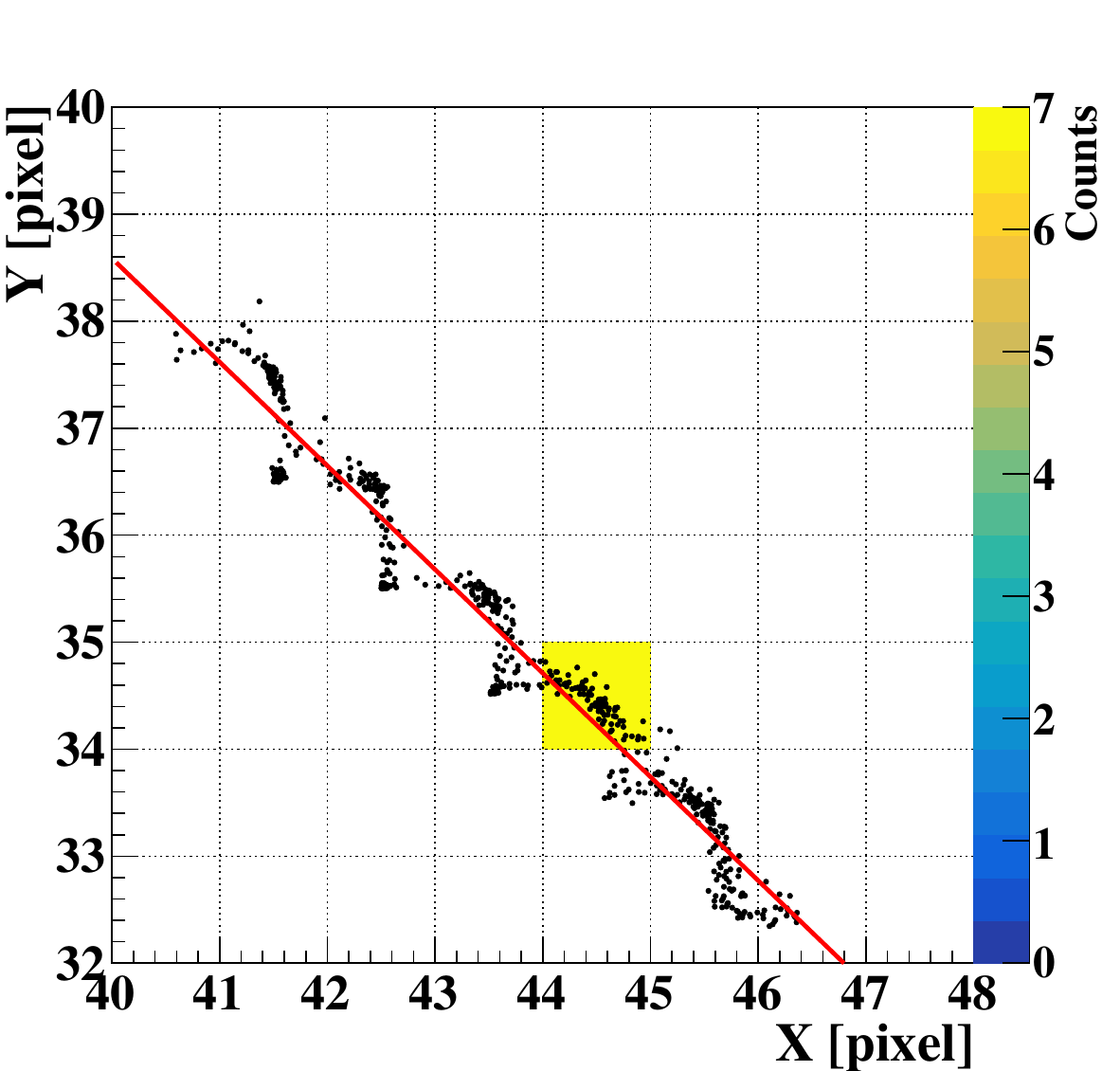}
\includegraphics[width=0.662\textwidth]{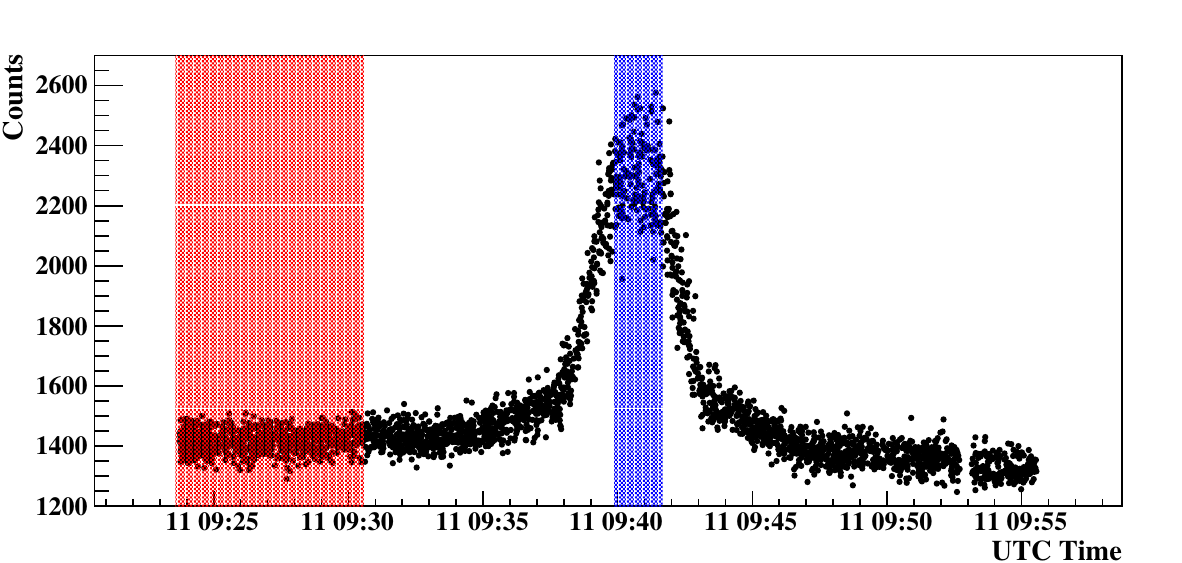}
\caption{Process of data analysis for a bright star HIP100453 with an apparent magnitude of 2.23$^m$. Black points in the left figure show positions of the star on the focal surface fitted with the Moffat function within one PMT. Then, the red line represents the star's track on the focal surface. At this stage, we can choose a pixel whose centre is placed at a distance not longer than 0.3 pixels from the track for further analysis. In the next step, we calculate the light curve of counts in a 3x3 pixels area around the chosen pixel. An example is presented on the right plot. Red and blue colours on the lightcurve represent the background and signal plus background ranges, respectively. The difference between average counts in those two periods gives the number of registered photons from the star.}
\label{DataAnalysis}
\end{figure}

\color{red}
}

\begin{figure}[ht]
\centering
\includegraphics[width=1.03\textwidth]{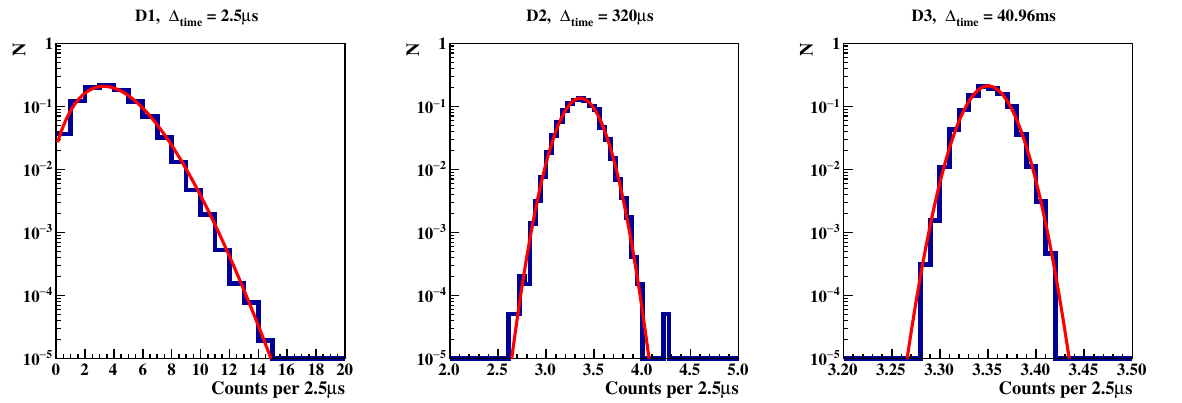}
\caption{Distribution of signal in one pixel of EUSO-TA2 detector measured during the same period in three time-scales: D1 with acquisition time of 2.5~$\mu$s - the same as for EUSO-TA (left), D2 with 320~$\mu$s (center) and D3 with 40.96~ms (right). All distributions are normalized. In D1 mode signal is well described with Poisson distribution, while in D2 and D3, we use the Gaussian one due to summed counts over a longer period. }
\label{PixelDist}
\end{figure}

The signal has been equalized between all pixels before the fitting procedure.
For equalization, we used analysis of signal distributions measured in background conditions before and/or after star transition. 
An example of signal distribution in one pixel with 2.5$\mu s$ GTU is presented in Figure~\ref{PixelDist} (left). 
After equalization, the signal in different pixels taken with the same background level is similar, but the efficiencies of various pixels can be different. 
Due to this, we introduce calibration constants $C(x,y)$ depending on the position on the focal surface. 
This factor describes the signal fraction registered in a $3\times3$ area around the pixel determined by~$x$ and~$y$ coordinates, including the PSF and an unknown part of the pixel's efficiency. 

\section{Stars as calibration sources}

In the analysis, we treat stars as point-like sources with known parameters. 
To express an expected number of registered photons $N(x,y, \Theta)$ coming from a star for a fixed observation angle $\Theta$ we use the following formula:

\begin{equation}
        N(x,y, \Theta) = C(x,y)\int_{0}^{\infty}T(\lambda, \Theta)\Phi(\lambda)P(\lambda)d\lambda,
\label{SignalFormula}
\end{equation}
where $T(\lambda, \Theta)$ denotes atmospheric transmittance, $\Phi(\lambda)$ is starlight flux at the top of the atmosphere, $P(\lambda)$ represents known detector parameters and $C(x,y)$ is a matrix of the wavelength-independent and position-dependent calibration constants.

Atmospheric transmission $T(\lambda, \Theta)$ for the defined observation angle is calculated using libRadTran package~\cite{gmd-9-1647-2016}, considering local atmospheric conditions at the BRM site. 
To calculate the number of photons from the star at the top of the atmosphere $\Phi(\lambda)$, we used 
Pickles stellar spectral flux library~\cite{Pickles:1998dm} renormalized for each star with its apparent magnitude. 
To recognize stars in the FoV, we used HIPPARCOS catalogue~\cite{Perryman:1997sa}. 
Known detector parameters $P(\lambda)$ include optical system efficiency and the size of the telescope lenses, quantum and collection efficiency of MAPMTs and dead time of the detector. In the first step of the analysis, data is corrected for the pile-up effect~\cite{Abdellaoui:2019qmg}.

\section{Results}
\begin{figure}[ht]
\centering
\includegraphics[width=1.01\textwidth]{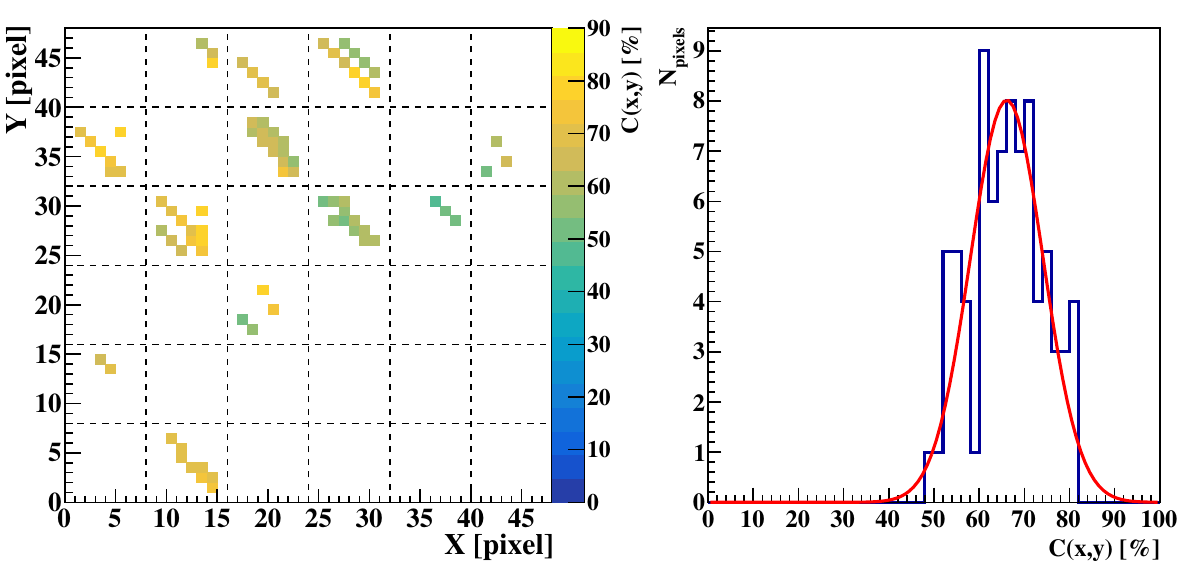}
\caption{Calculated calibration factors $C(x,y)$  at different positions on the focal surface (left) for stars measured in October 2015 and its distribution fitted with the Gaussian form (right). The mean value of (66.1 $\pm$ 8.1)$\%$ can be used to calculate the average efficiency of the detector.  }
\label{PsfFactor}
\end{figure}

\begin{figure}[ht]
\centering
\includegraphics[width=0.58\textwidth]{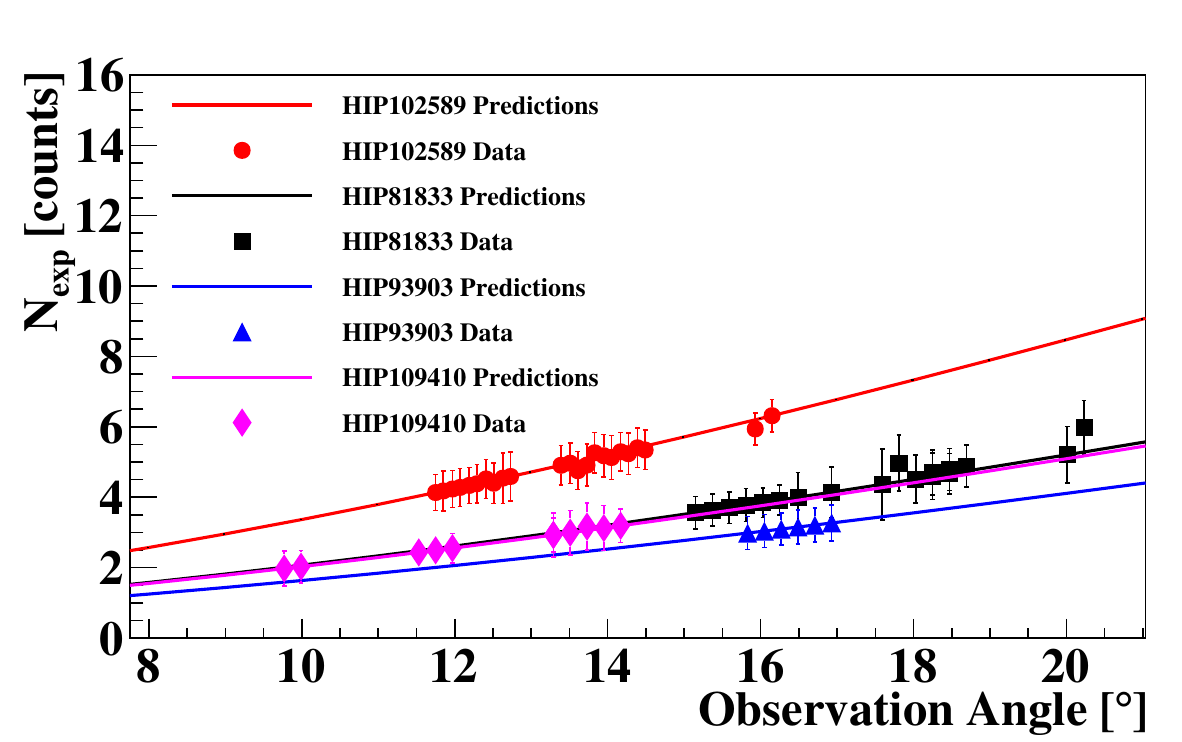}
\includegraphics[width=0.41\textwidth]{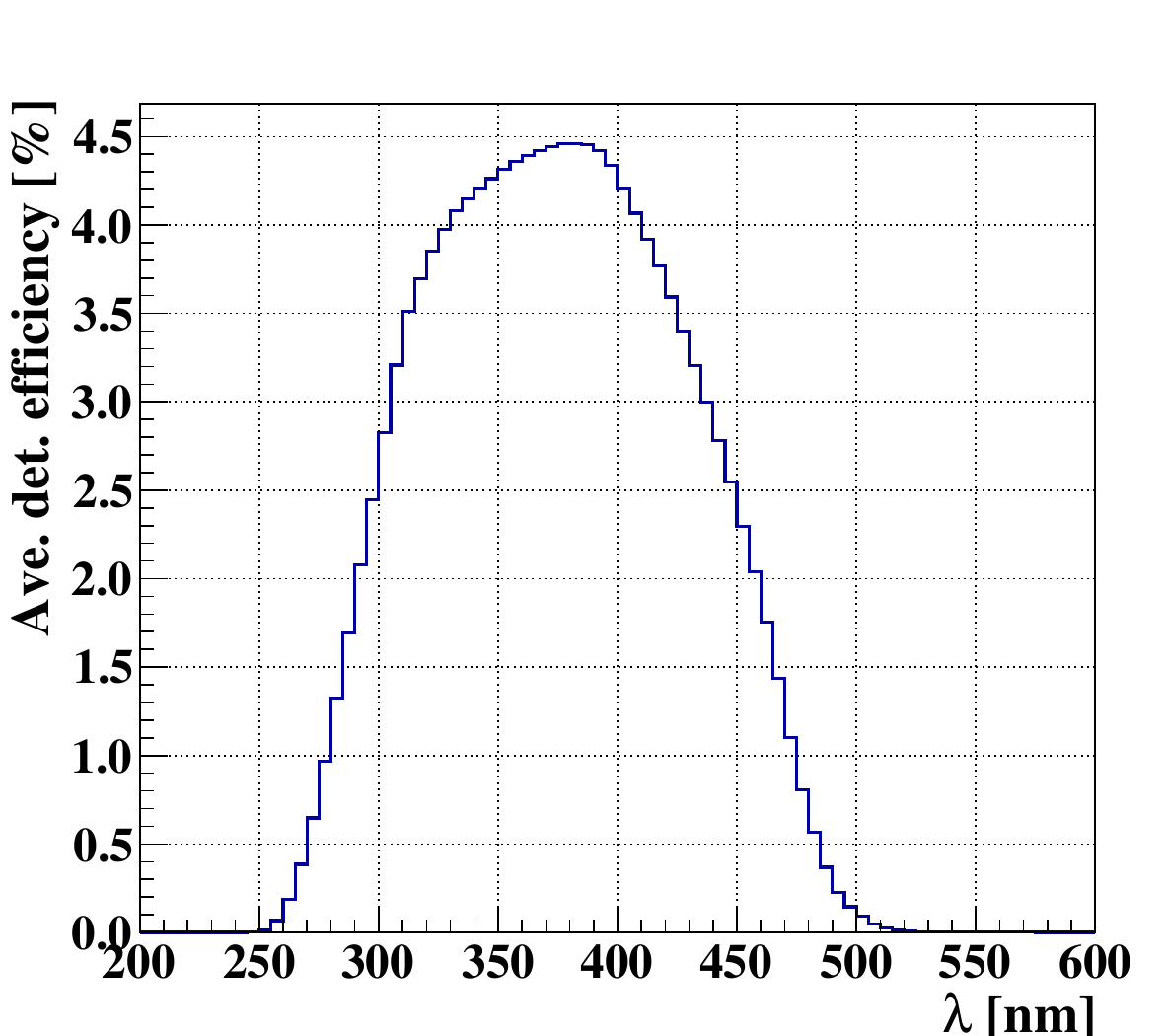}
\caption{The left plot shows a comparison between measured (points) and predicted (lines) signals for four stars with different intensities. Points have been corrected by obtaining $C(x,y)$ factors depending on the position on the focal surface during measurement. Lines represent expected signals assuming a PSF that allows to fit full signal in $3\times3$ pixels area. The right plot shows the calculated efficiency of the EUSO-TA detector, assuming an average value of $C(x,y)$ as 66.1$\%$. Calculated detector efficiency reaches the maximum of 4.46$\%$ at about 380nm.}
\label{DataVsPred}
\end{figure}

In this work, we analyzed 22 stars observed by the EUSO-TA telescope in 2015 and 2016. Additionally, we performed the first analysis of the data taken by EUSO-TA2 in 2022 and 2023. 
Calculated matrix of calibration constants 
$C(x,y)$ can differ for data sets taken during different EUSO-TA campaigns due to different optical axis positions, focusing, pixel gains and other parameters. 
Below, we present results for data taken by EUSO-TA in October 2015.
By fitting the measured signal to calculated predictions, we obtained the $C(x,y)$ matrix presented in figure~\ref{PsfFactor} (left). 
For 81 pixels, we obtain distribution ranging from 48$\%$ to 82$\%$ with an average of about 66$\%$ (figure~\ref{PsfFactor} right). 
Using calculated factors, a comparison between predicted and detected signals is presented in Figure~\ref{DataVsPred} (left) for four stars with different magnitudes, measured at different elevation angles. 
Predictions include all parameters from the right side of equation~(\ref{SignalFormula}) excluding $C(x,y)$, which is applied in the data, for each point separately, according to its position on the focal surface during observation. 
An average value of the calibration constant from Figure~\ref{PsfFactor} has been used to plot average detector efficiency as a function of wavelength presented in Figure~\ref{DataVsPred} right. 
The shape is determined mainly by BG3 filter characteristics and the quantum efficiency of MAPMTs. 
The detector reaches maximum efficiency of (4.46$\pm$0.55)\% at 380~nm. 
The average efficiency was estimated as (2.74$\pm$0.33)\% and (4.10$\pm$0.49)\% in the ranges 250-500 nm and 300-400 nm respectively.
{
\color{red}
}

\section{Conclusions}
{

In the presented analysis, we have considered signals from 22 stars detected by the EUSO-TA detector in 2015--2016 and a few stars observed by the upgraded version - EUSO-TA2 in 2022--2023.
We described the method of performing the absolute calibration of a multi-pixel detector using UV light from stars as a reference signal. 
Based on data taken in October 2015, we obtained calibration factors of 81 out of 2304 pixels on the focal surface. The EUSO-TA detector average efficiency has been calculated as (4.10$\pm$0.49)\% in the range 300 - 400 nm, consistent with previous estimations done with calibration lasers and LEDs.
The presented method provides valuable information about detector response during operation. 
Application of the presented analysis to EUSO-TA2 and EUSO-SPB2 data taken at the TA site will allow for estimating those detectors' efficiencies and help improve the method.
Such operations can be performed for future EUSO missions during field tests in Utah. 
The possibility of pointing the balloon-borne telescope towards stars during flight can also be considered.

}

{
\small{\bf Acknowledgments:}
This work was partially supported by Basic Science Interdisciplinary Research
Projects of RIKEN and JSPS KAKENHI Grant (JP17H02905, JP16H02426 and
JP16H16737), by the Italian Ministry of Foreign Affairs and International Cooperation, 
by the Italian Space Agency through the ASI INFN agreements Mini-EUSO n. 2016-1-U.0, EUSO-SPB1 n. 2017-8-H.0, OBP (n. 2020-26-Hh.0), EUSO-SPB2 n. 2021-8-HH.0 and by ASI INAF agreement n. 2017-14-H.O, by NASA awards and grants 11-APRA-0058, 16-APROBES16-0023, 17-APRA17-0066, NNX17AJ82G, NNX13AH54G, 80NSSC18K0246, 80NSSC18K0473, 80NSSC19K0626, 80\-NSSC\-18K\-0464 and 80NSSC22K1488 in the USA, Deutsches Zentrum f\"ur Luft- und Raumfahrt, by the French space agency
CNES, the Helmholtz Alliance for Astroparticle Physics, funded by the
Initiative and Networking Fund of the Helmholtz Association (Germany), by National Science Centre in Poland grant no 2017/27/B/ST9/02162 and
2020/37/B/ST9/01821. L. W. Piotrowski acknowledges financing by the Polish National Agency for Academic Exchange within Polish Returns Programme no. PPN/PPO/2020/1/00024/U/00001 and National Science Centre, Poland grant no. 2022/45/B/ST2/02889.
Russian team is supported by ROSCOSMOS, "KLYPVE" is included into the
Long-term program of Experiments on board the Russian Segment of the ISS.
Sweden is funded by the Olle Engkvist Byggm\"astare Foundation.
}

\bibliography{my-bib-database}

\newpage
{\Large\bf Full Authors list: The JEM-EUSO Collaboration\\}

\begin{sloppypar}
{\small \noindent
S.~Abe$^{ff}$, 
J.H.~Adams Jr.$^{ld}$, 
D.~Allard$^{cb}$,
P.~Alldredge$^{ld}$,
R.~Aloisio$^{ep}$,
L.~Anchordoqui$^{le}$,
A.~Anzalone$^{ed,eh}$, 
E.~Arnone$^{ek,el}$,
M.~Bagheri$^{lh}$,
B.~Baret$^{cb}$
% ...
}
\end{sloppypar}
\vspace*{.3cm}

{ \footnotesize
\noindent
% ...
$^{cb}$ Universit\'e de Paris, CNRS, AstroParticule et Cosmologie, F-75013 Paris, France\\
% ...
$^{ed}$ Istituto Nazionale di Fisica Nucleare - Sezione di Catania, Italy\\
% ...
$^{eh}$ INAF - Istituto di Astrofisica Spaziale e Fisica Cosmica di Palermo, Italy\\
% ...
$^{ek}$ Istituto Nazionale di Fisica Nucleare - Sezione di Torino, Italy\\
$^{el}$ Dipartimento di Fisica, Universit\`a di Torino, Italy\\
% ...
$^{ep}$ Gran Sasso Science Institute, L'Aquila, Italy\\
% ...
$^{ff}$ Nihon University Chiyoda, Tokyo, Japan\\ 
% ...
$^{ld}$ University of Alabama in Huntsville, Huntsville, AL, USA\\
$^{le}$ Lehman College, City University of New York (CUNY), NY, USA\\
% ...
$^{lh}$ Georgia Institute of Technology, USA\\
% ...
}



%
%
%

\end{document}